\documentclass[pre,a4paper,twocolumn,superscriptaddress,%
floatfix]{revtex4}

\usepackage[T1]{fontenc}
\usepackage[utf8]{inputenc}

\usepackage{graphicx}
\usepackage{color}
\usepackage{amsmath}
\usepackage{csquotes}

\begin{document}

\title{Nonequilibrium heat relation}

\author{Jean-Luc Garden}
\email{jean-luc.garden@neel.cnrs.fr}
\affiliation{Univ. Grenoble Alpes, CNRS, Grenoble INP, Institut N\'{E}EL,
38000 Grenoble, France}

\date{\today}

\begin{abstract}
The nonequilibrium work relation, or Jarzynski equality, establishes a statistical relationship between a series of nonequilibrium experiments on a system subjected to thermal fluctuations and a hypothetical experiment at thermodynamic equilibrium. In these experiments, the fluctuating quantity is the work exchanged between the system and its environment, while in the equilibrium scenario, the Helmholtz free energy difference between the system's initial and final states is determined. We inquire about the corresponding associated heat, the contribution of which, when added to the work, yields the change in internal energy. A new equality is presented for the random heat exchanged between the system and its thermal bath during the same protocol as the Jarzynski equality. Guidelines are provided for the experimental conditions required to measure such random heat.
\end{abstract}

\maketitle

\textit{Introduction}\textemdash  Stochastic thermodynamics studies thermodynamic behavior in macroscopic systems small enough to be sensitive to thermal fluctuations. In fields such as mesoscopic physics, molecular physics, and biophysics, measurable quantities (e.g., voltage, length, configuration,...) fluctuate due to the comparable scale of energy and momentum exchanges between the system and surrounding particles. At the molecular or atomic level, energy involved in measurements is particularly sensitive to thermal agitation for the same reason. Since physical measurements imply energetic interactions with the system, fluctuations can notably impact individual measurements, resulting in variability across experiments and yielding a distribution of observed values.\\
In recent decades, a series of theorems known as fluctuation theorems have been developed. For a comprehensive overview of stochastic thermodynamics, we refer to the following review articles \cite{EVANS1,MARCONI,JAR1,SEIFERT}.\\
Among the fluctuation theorems, the work fluctuation theorem, or the Jarzynski equality (JE), establishes a statistical relationship from a series of identical experiments, where random work $W$ is measured, and the Helmholtz free energy difference $\Delta F=F_{B}-F_{A}$ between the equilibrium final and initial state $B$ and $A$ respectively of the system \cite{JAR2,JAR3}:
\begin{equation}
  \label{eq:JE}
  \left\langle e^{-\beta W}\right\rangle=e^{-\beta \Delta F} .
\end{equation}
$W$ is the measured work done on the system during one experiment; $\beta =\frac{1}{k_{B}T}$ where $T$ is the temperature of the thermal bath inside which the system is immersed and $k_{B}$ the Boltzmann constant. The brackets indicate statistical averaging over the series of experiments following the work protocol (WP) underlying the JE. However, no study has established an equivalent theorem for the random heat exchanged during the process underlying the WP of the JE. The variation of the external control parameter during the WP induces heat transfer to or from the thermal bath. Building on Evans and coworkers fluctuation theorem for entropy production rates \cite{EVANS2}, Crooks proposed new transient fluctuation theorems for entropy production, work, and heat \cite{CROOKS1,CROOKS2}. Assuming microscopic reversibility of the equations of motion, he showed that the transient fluctuation theorem for work, upon integration, leads to the JE. For heat, for cyclic processes bringing the system in a  steady nonequilibrium state the following holds:
\begin{equation}
  \label{eq:CrooksQ}
  \lim_{{\Delta t \to +\infty}}\frac{P(+\beta Q)}{P(-\beta Q) }=e^{\beta Q} ,
\end{equation}
Here, $P$ represents the probability of the events in brackets. This heat fluctuation theorem holds in the long-time limit for cyclic steady-state processes where entropy production is primarily driven by heat transfer \cite{CROOKS2}. Van Zon and Cohen extended the fluctuation theorem for heat to stochastic transient processes \cite{VANZON1,VANZON2}:
\begin{equation}
  \label{eq:VANZONQ}
  \frac{P_{\tau}(Q_{\tau})}{P_{\tau}(-Q_{\tau})}\sim e^{\beta Q_{\tau}} .
\end{equation}
The symbol $\sim$ refers to the long-time limit for $\tau$ of steady-state processes, where Eq.~(\ref{eq:CrooksQ}) is recovered, while the equality holds for all times for transient processes only. According to the authors $Q$ represents the portion of work transferred to the surrounding fluid. A model is developped for $Q$ that allows the determination of its probability distribution function for a Brownian particle dragged through water by a moving potential, using the first law of thermodynamics and the known work and energy variations in this case. Baiesi and coworkers showed that, unlike work fluctuations, heat fluctuations exhibit an asymmetry, particularly for large values of $q$ (heat per unit time) \cite{BAIESI}. This results in a correction to the standard fluctuation theorem for heat (such as in Eq.~(\ref{eq:VANZONQ})), especially at high heat rates $q$. Hatano and Sasa further generalized the Jarzynski equality for steady-state nonequilibrium systems \cite{HATANO}:
\begin{equation}
  \label{eq:Hatano}
  \left\langle e^{-\beta Q_{ex}-\Delta \phi}\right\rangle=1 .
\end{equation}
Here, $Q_{ex}$ represents the \enquote{excess heat}, which reflects changes in the system's state space, and $\Delta \phi$ is related to the difference in the probability distribution function between the two steady states. In Ref. \cite{JAR4}, Jarzynski and W\'{o}jcik examined a fluctuation theorem for heat transfer between two systems prepared at two different heat bath temperatures:
\begin{equation}
  \label{eq:QJar1}
  \frac{P_{\tau}(+Q)}{P_{\tau}(-Q)}=e^{\Delta \beta Q}.
\end{equation}
Here, $\Delta \beta=\frac{1}{k_{B}T_{B}}-\frac{1}{k_{B}T_{A}}$ represents the difference between the inverse temperatures of the two baths, while the left-hand side denotes the ratio of the observed distribution of values of $Q$ exchanged over an ensemble of repetitions when the bodies are put in contact. In this context, since $\Delta \beta Q$ corresponds to the entropy production, the theorem addresses entropy production fluctuations due to heat transfer between two reservoirs. Noh and Park investigated heat fluctuations and proposed a fluctuation relation for heat \cite{NOH}:
\begin{equation}
  \label{eq:Noh}
  \frac{P_{h}(Q)}{P_{h}(-Q)}=e^{Q}/\Psi(Q) .
\end{equation}
In this context, $\Psi(Q)$ represents the correlation between energy change and heat, as both are linked through the conservation of energy. Recently, Pal and co-workers, along with Usha Devi and co-workers, examined the result of Eq.~(\ref{eq:QJar1}) \cite{PAL,USHA}. The former derived a modified expression of Eq.~(\ref{eq:QJar1}), and recovered the equality in Eq.~(\ref{eq:Noh}) for a single heat bath \cite{PAL}. The latter discussed Eq.~(\ref{eq:QJar1}) in the context of quantum Gaussian states in thermal equilibrium \cite{USHA}. More recently, Wu and An extended the result of Eq.~(\ref{eq:QJar1}) to systems beyond the weak system-bath coupling assumption, addressing strong coupling conditions via an effective system temperature \cite{WU}. This holds, as entropy production can always be defined based on an effective system temperature that differs from the bath temperature \cite{GARDEN0}.\\
In this letter, we present a nonequilibrium heat relation, which is an integrated form of a fluctuation theorem for heat. It applies under the same theoretical conditions as the JE. The random heat statistics is linked to the entropy difference between states $A$ and $B$, in the same way that random work statistics is connected to the free energy difference between the same equilibrium states. We also discuss the experimental conditions required to measure this random heat quantity during the WP.\\

\textit{Nonequilibrium heat relation}\textemdash Let us first consider macroscopic systems (hence the use of brackets for measured quantities). When work $\left\langle W \right\rangle$ is applied through an external control parameter $x$ (e.g., the elongation of a macromolecule) with a switching rate $r=dx/dt$ to drive the system from state $A$ to state $B$, we have:
\begin{align}
\label{eq:W}
 \left\langle W\right\rangle & =\left\langle \displaystyle \int_{x_{A}}^{x_{B}} \left(\partial F(x,T)/\partial x\right)_{T}dx\right\rangle \nonumber \\
 & =\left\langle \displaystyle \int_{t_{A}}^{t_{B}} \left(\partial F(x,T)/\partial x\right)_{T}rdt \right\rangle .
\end{align}
For an equilibrium transformation, since $1/T$ is an integrating factor of heat exchange we have:
\begin{align}
\label{eq:Qeq}
  \left\langle Q\right\rangle  = & \left\langle \int_{x_{A}}^{x_{B}} T\left(\partial S(x,T)/\partial x\right)_{T}dx \right\rangle \nonumber \\
  & + \left\langle\int_{T_{A}}^{T_{B}}T\left(\partial S(x,T)/\partial T \right)_{x}dT\right\rangle .
\end{align}
$S(x,T)$ is the system's entropy. Here, the variable $T$ is the temperature of the system, which can be different from bath temperature if thermal equilibrium is not fulfilled. However, even in an isothermal transformation where $T$ remains equal to the bath temperature, there is a contribution to the heat due to the change in $x$ as observed in Eq.~(\ref{eq:Qeq}). For non-equilibrium processes, Eq.~(\ref{eq:W}) remains valid while Eq.~(\ref{eq:Qeq}) does not, as $1/T$ is no longer an integrating factor for heat exchange. The measured force, $f=-\partial F(x,T)/\partial x$, however, is different than for equilibrium. Mechanical equilibrium is not fulfilled. Moreover, in this case an additional term must be considered in the heat exchange between the system and its surroundings due to internal relaxation processes within the system itself \cite{GARDEN1}.  Assuming for the sake of simplicity that irreversible processes are characterized by one single parameter $\xi$, all state functions $F$, $S$ and $U$ (the internal energy), depend on this additional state variable of the system \cite{PRIGO}. $\xi$ represents the matter repartition in the system, and the corresponding measured work $\left\langle W\right\rangle$ is now dependent on it. This dependency is observed in the work dependency on the switching rate $r$ observed in experiments following the WP \cite{LIP}. The WP is illustrated in Fig. 1, where a biomolecule is stretched from an initial equilibrium state $A$ with free energy $F_{A}$ and coordinate $x_{A}$ to a final state $B$ with free energy $F_{B}$ and coordinate $x_{B}$ over a time $\Delta t$. One end of the molecule is attached to a bead held by a pipette, while the other is connected to a bead in an optical trap that measures force along the transformation from $A$ to $B$. The $F(x)$ graph displays four trajectories corresponding to different switching rates $r=dx/dt$, which relate to the internal coordinate $\xi$. The dotted straight line represents the equilibrium path for an infinite $\Delta t$, where equilibrium is maintained. Along the trajectory $\xi_{1}(x)$ (two of them are visible), work $W$, and heat $Q$, are exchanged with the surroundings. The shaded area around the two visible paths $\xi_{1}(x)$ illustrates the uncertainty in measured quantities due to fluctuations.
\begin{figure}[h]
  \centering
  \includegraphics[width=8.6 cm]{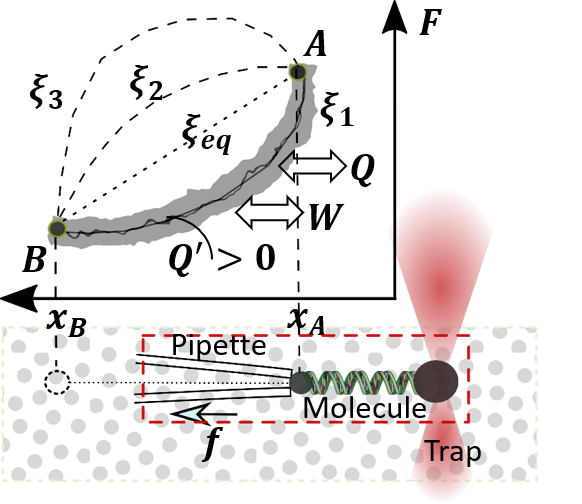}
  \caption{The work protocol is illustrated with a biomolecule stretched by a pipette from the position $x_{A}$ to $x_{B}$, while the force is measured with an optical trap.  Four different trajectories are represented in a $F(x)$ graph, each of them being $\xi $-dependent. Along the trajectory $\xi_{1}(x)$, the stochastic quantities $W$, $Q$ and $Q'$ are represented. The red dashed rectangle refers to the new system depicted in Fig. 2}
  \label{fig:Fig1}
\end{figure}
For nonequilibrium trajectories, Eq.~(\ref{eq:W}) remains valid, but $F = F(x, T, \xi)$, and the partial derivative of the free energy (force) is taken now with respect to constant $T$ and $\xi$. The system's dissipative processes due to internal relaxation $d\xi/dt$ in the time interval $dt$ (during the change of the macromolecule conformation as in \cite{LIP} for example), are described by the following expression:
\begin{equation}
\label{eq:Q'}
  \left\langle \delta Q'\right\rangle= \left\langle T\delta_{i}S\right\rangle=\left\langle Ad\xi \right\rangle \geq\ 0.
\end{equation}
$\left\langle Q'\right\rangle=\left\langle\int_{A}^{B} \delta Q' \right\rangle  $ represents the uncompensated heat of Clausius \cite{DEDONDER}. $\left\langle Q'\right\rangle$ is the heat generated inside the system that has not had sufficient time to be exchanged with the thermal bath along the trajectory from $A$ to $B$ (Cf. Fig 1 where $Q'$ is generated internally to the molecule). $A=-\left(\partial F(x,T,\xi)/\partial \xi \right)_{x,T}$ denotes the affinity of the irreversible process \cite{DEDONDER}. $\left\langle \Delta_{i}S\right\rangle=\left\langle \int_{A}^{B} \delta_{i}S \right\rangle$ represents, in addition to the entropy exchanged with the surroundings (i.e. heat), the positive entropy produced along the path between the two equilibrium states $A$ and $B$. As discussed in the introduction, when the system becomes \enquote{small}, not only $W$, but also $Q$ and $Q'$ become stochastic quantities (this point is illustrated in Fig. 1 by the shadded area around the $\xi_{1}(x)$ trajectories). From this point on, the brackets are omitted to treat these quantities as random. During the WP of JE, $W$ and $Q$ are connected through the first law of thermodynamics (for closed systems):
\begin{equation}
\label{eq:FirstLaw}
  \Delta U=U_{B}-U_{A}= Q + W  .
\end{equation}
This equality holds exactly, regardless of whether the transformation is at equilibrium or not. Since internal energy is a state function, $\Delta U$ does not depend on the path connecting equilibrium states $A$ and $B$, and it remains constant, independent of switching rates, or the value of $\xi$, and independent of fluctuations during the WP. The same applies to all state functions (particularly $F$ and $S$). Consequently, we immediately have:
\begin{equation}
\label{eq:FLUCU}
   \left\langle e^{-\beta Q}\times e^{-\beta W}\right\rangle=e^{-\beta \Delta U} .
\end{equation}
This is the first nonequilibrium relation presented in the paper, which may initially appear trivial. With the use of the uncompensated heat of Clausius $Q'$ defined above, the second law is expressed as follows during the WP:
\begin{equation}
\label{eq:EqFond1}
  Q'=\int_{S_{A}}^{S_{B}}TdS-Q= W-\Delta F-\int_{T_{A}}^{T_{B}}SdT.
\end{equation}
The Helmholtz free energy is related to the internal energy by $F = U - TS$. $Q'$, associated with entropy production, is a stochastic variable that can be either positive or negative, depending on the fluctuations during the protocol. Given that the WP occurs under isothermal conditions, as is the case, for instance, with a macromolecule immersed in an aqueous solvent (e.g. \cite{LIP}), the above expression simplifies to:
\begin{equation}
\label{eq:EqFond1bis}
  Q'=T\Delta S-Q= W-\Delta F.
\end{equation}
By taking the statistical average of each term above, the two equalities involving $Q'$ are both expressions of the second law of thermodynamics, given that $\left\langle Q' \right\rangle \geq 0$. We observe that there are two equalities involving the same $Q'$, each expressed as the difference between a random variable and a constant: one involving $W$ and the other $Q$. Consequently, multiplying these equalities by $-\beta$, taking the exponential, and performing the statistical averaging yields:
\begin{equation}
\label{eq:EqFond2}
\left\langle e^{-\beta Q'}\right\rangle=e^{-\frac{\Delta S}{k_{B}}}\times \left\langle e^{\beta Q}\right\rangle=\left\langle e^{-\beta W}\right\rangle\times e^{\beta \Delta F}.
\end{equation}
From the JE (Eq.~(\ref{eq:JE})), the right term in Eq.~(\ref{eq:EqFond2}) equals unity, resulting in the nonequilibrium heat relation:
\begin{equation}
\label{eq:Heat}
  \left\langle e^{\beta Q}\right\rangle=e^{\frac{\Delta S}{k_{B}}}.
\end{equation}
This is the central result of the paper. A series of measurements of $Q$ following the experimental protocol (WP) underlying the JE directly provides the entropy difference between states $A$ and $B$. This entropy difference corresponds to the maximum heat, $\left\langle Q_{max} \right\rangle$, that can be achieved for a system remaining in equilibrium between $A$ and $B$. We have $\left\langle Q_{max} \right\rangle=T\Delta S$ just as $\left\langle W_{min} \right\rangle=\Delta F $ for isothermal process. By applying Jensen's inequality to this equation, we obtain the Clausius inequality, $\Delta S \geq \frac{\left\langle Q \right\rangle}{T}$. From Eq.~(\ref{eq:EqFond2}), Crooks' fluctuation theorem for entropy production is recovered \cite{CROOKS2}:
\begin{equation}
  \label{eq:DiS}
  \left\langle e^{-\beta Q'}\right\rangle=\left\langle e^{-\frac{\Delta_{i} S}{k_{B}}}\right\rangle=1 .
\end{equation}
In fact, the reasoning can be reversed. From the works of Evans and colleagues and Crooks, we know that Eq.~(\ref{eq:DiS}) holds exactly for all random entropy produced during a transformation from an equilibrium state $A$ to another equilibrium state $B$ over a time scale $\tau$ \cite{EVANS2, CROOKS1, CROOKS2}. If the transformation takes place under isothermal conditions, then from the two equalities in Eq.~(\ref{eq:EqFond2}), both the JE and the nonequilibrium heat relation (Eq.~(\ref{eq:Heat})) can be derived at the same level. A similar expression to Eq.~(\ref{eq:Heat}) for the entropy difference has been derived by Adib for large systems undergoing isoenergetic processes, where $U_{B}=U_{A}$ and thus $W=Q$ in this context \cite{ADIB}.\\
To conclude this section, another consequence of the JE and Eq.~(\ref{eq:Heat}) leads to the following nonequilibrium relation:
\begin{equation}
\label{eq:U}
  \frac{\left\langle e^{-\beta W}\right\rangle}{\left\langle e^{\beta Q}\right\rangle}=e^{-\beta \Delta U}.
\end{equation}
Measurements of the random work and random heat in a series of experiments conducted out of equilibrium during the WP provide direct access to the internal energy difference between the two equilibrium states $A$ and $B$. Ultimately, equating Eq.~(\ref{eq:U}) and Eq.~(\ref{eq:FLUCU}) gives:
\begin{equation}
\label{eq:WQ}
 \left\langle \frac{e^{-\beta W}}{e^{\beta Q}} \right\rangle=\frac{\left\langle e^{-\beta W}\right\rangle}{\left\langle e^{\beta Q}\right\rangle},
\end{equation}
which is also a remarkable relation.\\

\textit{Proposal of experiment to measure Q}\textemdash In this section, we refer to experiments conducted on biomolecules immersed in a solvent like for exemple in Refs. \cite{LIP,COLIN,HUGUET,JUNIER,ALEMANY,CAMUNAS,RISSONE}. However, other types of experiments on physical small objects, such as those involving coloidal particles in optical traps \cite{WANG,CARBERRY}, single-electron transistors \cite{MAILLET}, or oscillator in double-well potential \cite{BELLON}, which have validated the JE and showed unintuitive probabilities behaviours, are also applicable. Since the relations in the preceding sections are valid only under isothermal conditions, we propose measuring heat exchanges between the system and its surroundings in an isothermal environment. This approach falls within the field of isothermal calorimetry \cite{BROWN}. One method involves using a material with first order phase transition with a known latent heat that is in good thermal contact with the system during the WP. Under these circumstances, heat from the system is transferred isothermally to the material which changes the ratio of its two phases during a first order phase transition. This method dates back to the first calorimetric measurements by Lavoisier and De Laplace \cite{LAVOISIER}. Here, we propose a second approach: employing a thermocouple to measure differential temperature changes, along with a second thermoelement acting as a compensating power by Peltier effect to maintain isothermal conditions (Cf. Fig. 2). Unlike the setup in Fig. 1, where the system comprises the molecule and the thermal bath (the surrounding solvent), in Fig. 2, the macromolecule is immersed in a small volume of solvent, with the system defined as the combination of both the molecule and the solvent (Cf. the red dashed rectangle in Fig. 1).  
\begin{figure}[h]
  \centering
  \includegraphics[width=8.6 cm]{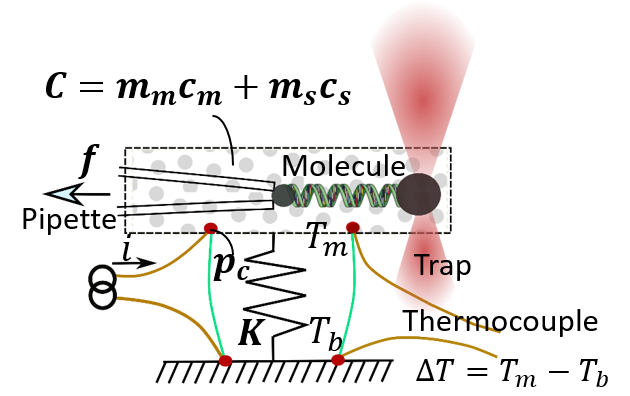}
  \caption{The biomolecule and a small solvent volume (heat capacity $C_{m}+C_{s}$) are thermally insulated from a bath at temperature $T_{b}$ by a thermal link with conductance $K$. A thermocouple measures the differential temperature $T_{m}-T_{b}$. A additional thermocouple crossed by a current $i$ compensates the system's power by Peltier power $p_{c}$ during the same work protocol as in Fig. 1 to determine the heat $Q$ exchanged with the thermal bath.}
  \label{fig:Fig2}
\end{figure}
The crucial condition for being able to measure the heat is that the solvent volume is also small (as well as the two thermoelements depicted in Fig. 2, and all other addenda that we consider of negligeable masses). The heat capacity of this system is $C = m_{m}c_{m} + m_{s}c_{s}$, where $c_{m}$, $m_{m}$ and $c_{s}$, $m_{s}$  are the specific heat and mass of the macromolecule and solvent respectively. The molecule and solvent are of strong thermal coupling, like it is the case in WP, so that the temperature is homogeneous throughout the system when measured with one bound of the thermocouple placed at any location (Cf. Fig. 2). Unlike in the experiments cited earlier, the solvent does not act as a thermal bath for the macromolecule because $m_{s}c_{s} \sim m_{m}c_{m}$ or $m_{s}c_{s} < m_{m}c_{m}$. To measure heat exchanges, the \enquote{new system} must be thermally decoupled from its thermal bath. It is thermally decoupled from the external thermal bath through a heat exchange coefficient $K$, as indicated by small thermal links (schematically represented by a spring in Fig. 2). $K$ should, for instance, be defined using small arms of micro-suspended membranes (Cf. Fig. 1 and Fig. 5 in Ref.~\cite{GARDEN2} for example). The thermal bath (represented by a mass symbol in Fig. 2), must be represented by a copper plate or all other highly conductive materials of very high mass (and specific heat) with the other bound of the thermocouple beneath it. Under these conditions, it becomes possible to measure the temperature difference between the system and its thermal bath during the WP, while using the feedback control loop of the second thermocouple to adjust the power compensation by Peltier effect. This allows for measuring the power required to keep the differential temperature at zero at all times, and thus to remain in isothermal conditions during the WP. The main experimental challenge in such nanocalorimetric measurements is the ability to thermally isolate a small molecule with solvent from its surroundings, accurately measure its local temperature and the corresponding heat of compensation, and maintain the capacity to perform work on the system by varying an external control parameter like in the WP. While this poses significant difficulties for biomolecules, it may be achievable in nano-systems where work can be exchanged electrically, and the temperature can be measured via higly sensitive hot-electron nanoscale quantum calorimeters as an example \cite{PEKOLA1,PEKOLA2} (see also Ref. \cite{BLICKLE} where Blickle and co-workers measured heat distribution from a colloidal particle in a time-dependent nonharmonic potential).  If the feedback power compensation loop is not active during the WP between equilibrium states A and B, the total amount of heat exchanged is given by the following equation:
\begin{equation}
\label{eq:Calo1}
Q=\displaystyle \int_{t_{A}}^{t_{B}} \frac{\delta Q}{dt}dt=\displaystyle \int_{t_{A}}^{t_{B}} C\frac{dT_{m}}{dt}dt+\displaystyle \int_{t_{A}}^{t_{B}} K\Delta Tdt  ,
\end{equation}
where $\Delta T=T_{m}-T_{b}$ with $T_{m}$ the temperature of the molecule + solvant, and $T_{b}$ is the bath temperature. With the compensating feedback system from the second thermocouple, $T_{m}$ is enslaved to $T_{b}$ ensuring that $Q_{c}=- Q$, where $Q_{c}=\int_{A}^{B} p_{c}dt$ represents the total Peltier heat of the compensating system during the transformation from state $A$ to state $B$. The measured heat of isothermal compensation is a stochastic quantity, similar to the heat generated in the WP. Finally, like $W$, $Q$ in Eq.~(\ref{eq:Calo1}) is a measurable random quantity, governed by the nonequilibrium heat relation in Eq~(\ref{eq:Heat}).\\

\textit{Conclusion}\textemdash In this letter, we derived a nonequilibrium heat relation based on the uncompensated heat of Clausius, establishing a connection between the stochastic heat measured in a series of nonequilibrium experiments and the entropy difference between two equilibrium states, $A$ and $B$. This entropy difference corresponds to the maximum heat extractable in a hypothetical equilibrium process conducted over an infinite timescale. The proposed relation complements Jarzynski's nonequilibrium work relation, which connects the nonequilibrium work to the free energy difference between the states $A$ and $B$. From Jarzynski equality and the nonequilibrium heat relation, additional nonequilibrium equalities can be derived, linking the nonequilibrium work and heat to the internal energy difference between $A$ and $B$. We also discussed the experimental conditions required to measure the random heat during the work protocol in isothermal conditions. For heat measurements, the relevant fluctuating variable is the system temperature, analogous to the force in work measurements \cite{GARDEN3}.

\begin{acknowledgments}
The author thanks B. Brisuda, E. Collin and O. Bourgeois (Institut Néel, Grenoble, France) for a careful reading of the manuscript and judicious comments. We thank Servier Medical Art (https://smart.servier.com/) for the molecule drawing in figures.
\end{acknowledgments}

\end{document}